\documentclass[12pt,preprint]{aastex}
\begin{document}
\title{MAGIC upper limits on the very high energy emission from GRBs}

%
\author{
 J.~Albert\altaffilmark{a}, 
 E.~Aliu\altaffilmark{b}, 
 H.~Anderhub\altaffilmark{c}, 
 P.~Antoranz\altaffilmark{d}, 
 A.~Armada\altaffilmark{b}, 
 C.~Baixeras\altaffilmark{e}, 
 J.~A.~Barrio\altaffilmark{d},
 H.~Bartko\altaffilmark{f}, 
 D.~Bastieri\altaffilmark{g}, 
 J.~Becker\altaffilmark{h}
% ...
}
 \altaffiltext{a} {Universit\"at W\"urzburg, D-97074 W\"urzburg, Germany}
 \altaffiltext{b} {Institut de F\'\i sica d'Altes Energies, Edifici Cn., E-08193 Bellaterra (Barcelona), Spain}
 \altaffiltext{c} {ETH Zurich, CH-8093 Switzerland}
 \altaffiltext{d} {Universidad Complutense, E-28040 Madrid, Spain}
 \altaffiltext{e} {Universitat Aut\`onoma de Barcelona, E-08193 Bellaterra, Spain}
 \altaffiltext{f} {Max-Planck-Institut f\"ur Physik, D-80805 M\"unchen, Germany}
 \altaffiltext{g} {Universit\`a di Padova and INFN, I-35131 Padova, Italy}  
 \altaffiltext{h} {Universit\"at Dortmund, D-44227 Dortmund, Germany}
% ...

\begin{abstract}

During its first data cycle, between 2005 and the beginning of year
2006, the fast repositioning system of the MAGIC Telescope allowed the
observation of nine different GRBs as possible sources of Very High
Energy (VHE) $\gamma$-rays.
These observations were triggered by alerts from
Swift, \mbox{HETE-II}, and Integral; they
started as fast as possible after the alerts and lasted for several minutes,
with an energy threshold varying between 80 and 200~GeV, depending upon the
zenith angle of the burst. No evidence for gamma signals was found,
and upper limits for the flux were
derived for all events using the standard analysis chain of MAGIC.
For the bursts with measured redshift, the upper limits are compatible
with a power law extrapolation,
when the intrinsic fluxes are evaluated taking into account
the attenuation due to the scattering in the Metagalactic Radiation Field (MRF).

\end{abstract}


\section{Introduction}

The physical origin of the enigmatic Gamma-Ray Bursts is still under debate today, 
40 years after their discovery \citep[see][for a recent review]{MeszarosROP}. 
The possible detection of radiation in the Very High Energy region
(extending between few tens of GeV and few tens of TeV) will lead
to a deeper understanding 
of the acceleration mechanisms and the emission processes from gamma-ray bursts. 
The $\gamma$-ray emission observed by the Energetic Gamma-Ray Experiment Telescope (EGRET) in some case extends up to the VHE band  \citep{Hurley, Dingus, Gonzalez},  
favouring the hypothesis of
a highly relativistic source of non-thermal radiation situated in an
optically thin region \citep{piran99};
more insight, however, can be gained by a clear signal detection in the VHE region,
or the evaluation of stringent upper limit in this energy band.

Several observations of GRBs at energies above 100 GeV
have been attempted  
\citep{HEGRA,Asgamma},
without showing any indication of a signal. This is due to
relatively low sensitivity, as in satellite-borne detectors, or to
high energy thresholds, as in the previous generation of  Cherenkov
telescopes or in particle detector arrays.
Only few tentative detections of radiation above 0.1~TeV 
were reported by MILAGRITO for GRB~970417a \citep{milagrito} 
and by the GRAND array on GRB~971110~\citep{poirer}.   

Upper limits on the prompt or delayed emission of GRBs were also set by Whipple \citep{whipple, whipple2}, 
MILAGRO \citep{milagro, milagro1, milagro2, GCN6},
STACEE \citep{STACEE}
and HEGRA-Airobicc \citep{padilla}.

Imaging Atmospheric Cherenkov Telescopes (IACT) of the latest generation
achieve a better flux 
sensitivity and a lower energy threshold, and thus are better suited to
detect VHE $\gamma$-rays;
on the other hand, their small fields of view permit unguided observations only
by virtue of serendipitous
detection, and they have to rely on an external trigger, such as that
provided by automated satellite link to the GRB Coordinates
Network (GCN),
which broadcasts the coordinates of events
triggered and selected by dedicated satellite detectors.

Among the new  Cherenkov telescopes, MAGIC \citep{MAGIC}
is best suited for the detection of the prompt emission
of GRBs, because of its low energy threshold, large effective
area and, in particular, its capability for fast slewing \citep{drive}. 
The low trigger threshold, currently 50~GeV at zenith,
should allow the observation of GRBs even at large redshift,
as lower energy radiation can effectively reach Earth without
interacting much with the MRF. Moreover, in its fast-slewing
mode, MAGIC can be repositioned within 30~s to any position
on the sky; in case of a target-of-opportunity alert by
GCN, an automated procedure takes only few seconds to terminate
any pending observation, validate the incoming signal,
and start slewing toward the GRB position. Up to now,
the current maximal repositioning time is $\sim100$~s.
In two cases, this allowed to put upper limits on the GRB flux
even during the prompt emission \citep{MAGIC_GCN, albert06, Morris2006}.

The detection of VHE radiation from the GRB is important for comparing different 
theoretical models. The emission in the GeV-TeV range in the prompt and delayed phase 
is predicted by several authors \citep[see][for a detailed analysis]{ZHANG,ASAF2, RAZZAQUE}. 
Possible processes comprise leptonic and hadronic models: inverse-Compton (IC) 
scattering by electrons in internal
\citep{PAPATHANASSIOU94,PILLA} or external shocks \citep{MESZAROS94}, 
IC in the afterglow shocks \citep{DCM, ZHANG, DERISHEV, WANG}, 
IC by electrons responsible for optical flashes   
\citep{BELOBORODOV}, and pure electron-synchrotron \citep{ZHANG}; 
proton-synchrotron emission \citep{TOTANI}, 
photon-pion production \citep{WAXMAN, BOETTCHER,CHIANG, LI, FRAGILE}, 
and neutron cascades \citep{BM, DRV, ROSSI}. 

During the early afterglow phase, the recent observations by the
Swift satellite of X-ray flares lasting $10^3\div10^5$~s \citep{Burrows} 
suggested an extended activity in the central engine of the GRB, and thus emission from
late internal shocks \citep{Kobayashi2007} or from refreshed shocks due to
energy injections at later time \citep{GUETTA2006}. 
In some cases, the energy release of these flares can be of the same order
of magnitude of the energy release in the prompt phase, as reported for GRB~050502b.
Possibility of correlated $\gamma$-ray emission
extending into the GeV-TeV region is predicted as well,
where the corresponding VHE flares are predicted to originate
from IC scattered photons in the forward shock \citep{XIANG-YU}.
Thus, observation of the delayed activity is of particular interest being in most
cases not constrained by the alerting and slewing time, and being still connected
to the investigation of the central engine dynamics.


Measurements in this energy range can be used to test all these competing models.
However, as most of the observed GRBs occur at large redshift, strong attenuation of the 
VHE $\gamma$-ray flux is expected, as a result of the interaction with low energy photons
of the MRF \citep{Nikishov,deJager}.
The knowledge of the redshift, therefore, is important for a precise 
interpretation \citep{MANNHEIM}.

In this article, we report on the analysis of data collected on several GRBs followed by MAGIC during 
their prompt emission and early afterglow phases.

\section{Gamma-ray analysis with the MAGIC telescope}

The Major Atmospheric Gamma Imaging Cherenkov (MAGIC) telescope
\citep{MAGIC}, located  on the Canary Island of
La~Palma (2200 m asl, 28$^\circ$45$'$~N, 17$^\circ$54$'$~W), completed its
commissioning phase in early fall 2004. 
MAGIC is currently the largest IACT,
with a 17~m diameter tessellated reflector dish consisting of 964
0.5~$\times$~0.5~m$^2$ diamond-milled aluminium mirrors. In its
current configuration, the MAGIC photo-multiplier camera has a 
trigger region of $2.0^\circ$ 
diameter~\citep{cortina}, and a trigger collection area for
$\gamma$-rays of the order of $10^5$~m$^2$, which increases further with the zenith
angle of observation. Presently, the accessible trigger energy range spans from
50-60~GeV (at small zenith angles) to tens of TeV.  The MAGIC telescope is 
focused to 10 km distance 
-- the most likely height at which a 50~GeV $\gamma$-ray shower has its maximum.
The accuracy in reconstructing the direction of incoming
$\gamma$-rays, the
point spread function (PSF), is about $0.1^\circ$, slightly
depending on the analysis.

The reconstructed signals are calibrated~\citep{callisto}, and then cleaned of 
spurious backgrounds 
from the light of the night sky using two different image cleaning procedures: 
one algorithm requiring signal exceeding fixed reference levels, and a second algorithm 
employing additionally the reconstructed information of the arrival time~\citep{gaug}.
Non-physical background images are eliminated (e.g.\ car flashes having triggered 
the readout). Events are processed
by means of the MAGIC standard analysis software~\citep{mars},
using the standard Hillas analysis~\citep{Hillas,Fegan}. 
Gamma/hadron separation is performed by
means of the Random Forest (RF) method \citep{Breiman}, a classification
method that combines several parameters describing
the shape of the image into a new parameter called \emph{hadronness} \citep{Heng},
the final $\gamma$/hadron discriminator in our analysis.
Monte Carlo samples are used to optimize, as a function of energy, the cuts in hadronness.
The energy of the $\gamma$-ray is also estimated using a RF approach,
yielding a resolution of $\sim30\%$ at $200\:\mathrm{GeV}$. 
The parameter \emph{alpha} of the Hillas analysis, which is
related to the direction of the incoming shower,
is not included in the calculation of hadronness, as 
it is used separately to evaluate the significance of a signal.
If the telescope is directed at a point-like $\gamma$-ray source, 
as a GRB is expected to be,
the alpha-distribution of collected photons should peak at $0\degr$,
while it is uniform for isotropic background showers.

\section{Blind Test with Crab Nebula \label{Crab}}

On 2005 October 11 at 02:17:37~UT, 
the Integral satellite announced GRB~051011\footnote{IBAS alert nr. 2673, see at \url{http://ibas.mi.iasf.cnr.it/}} at the 
position J2000 \mbox{R.A. = $5^\mathrm{h}\: 34^\mathrm{m}\: 47^\mathrm{s}$}, 
\mbox{decl. = $+21^\circ \: 54' \: 39''$}.
A few hours later, Integral sent a new GCN
\citep{grb051011conf} stating that
GRB~051011 was in fact the Crab Nebula.
Thus, in a blind test, we acquired 2814 seconds of events coming from
the Crab Nebula, the standard source
of $\gamma$-rays at VHE energies.
The analysis yielded a $14\sigma$ signal above 350 GeV \citep{scapin},
showing that MAGIC can observe, at $5\sigma$ level, spectra of 5 Crab
Units (1 C.U. $= 6.57\times 10^{-10}\; \mathrm{cm^{-2}\; s^{-1}}$ above 100~GeV) 
of intensity in 40~s, if above 300~GeV, and in 90~s if below 300~GeV.

\section{GRBs observed by MAGIC during its first observation cycle}

An automatic alert system  has been operational from July 15$^\mathrm{th}$, 2004.
Since then, about 200 GRBs were detected by HETE-II, Integral and Swift, out of which
about 100 contained GRB coordinates. Time delays to the onset of the burst
were of the order of several seconds to tens of minutes.
During the first MAGIC data cycle, between April 2005 and March 2006,
9 GRBs were observed by MAGIC during the prompt or the early afterglow
emission phase, as listed in Table~\ref{tab:list}.
In two cases the prompt alerting by the GCN and the fast reaction of the MAGIC telescope allowed
to take data not only on the early afterglow, but also on part of the prompt emission of the burst.
These two bursts are GRB~0507013a and GRB~050904 and will be considered separately

\begin{table}[htb]
\begin{center}
\begin{tabular}{r|l|c|r|r|r|r|r|c}
   & Burst  & Satellite & $T_{0} $ [UT] & $\Delta T_\mathrm{alert}$ & $\Delta T_\mathrm{start}$ & $t_\mathrm{slewing}$ & Data & zenith angle \\
\hline
1. & GRB~050421 & Swift  & 04:11:52 & 58  s  & 108  s  & 26  s   & 75 min  & $\sim 52^\circ$  \\
2. & GRB~050502a   & Integral  & 02:13:57 & 39  s  & 689  s    &  223  s    & 87 min  & $\sim 30^\circ$  \\
3. & GRB~050505   & Swift  & 23:22:21   & 540 s  & 717  s   & 90  s     & 101 min    & $\sim 49^\circ$  \\
4. & GRB~050509a & Swift  & 01:46:29    & 16  s  & 131 s  & 108  s    & 119 min     & $\sim 58^\circ$  \\
5. & GRB~050713a & Swift  & 04:29:02 & 13 s & 40  s  & 17  s     & 37 min & $\sim 49^\circ$ \\
6. & GRB~050904    & Swift  & 01:51:44 & 82  s   & 145  s      & 54  s    & 147 min     & $\sim 24^\circ$ \\
7. & GRB~060121    & HETE-II  & 22:24:54  & 15  s  & 583  s    & --         & 53 min   & $\sim 48^\circ$ \\
8. & GRB~060203  & Swift    & 23:55:35  & 171  s     & 268  s    & 84  s  & 43 min   & $\sim 44^\circ$ \\
9. & GRB~060206  & Swift    & 04:46:53   & 16  s  & 59  s   & 35  s  & 49 min  & $\sim 13^\circ$ \\
\end{tabular}\\
\end{center}
\caption{\label{tab:list} Summary of GRBs observed by MAGIC from April 2005 to March 2006. 
$\Delta T_\mathrm{alert}$ stands here for the time delay after $T_0$ until 
the burst coordinates were received 
from the GCN. $\Delta T_\mathrm{start}$ is the total time delay 
before the observation could be started, of which
$t_\mathrm{slewing}$ is the time lost for repositioning the telescope. Data column shows the total amount of data taken.} 
\end{table}

\subsection{The properties of observed GRBs}

Table~\ref{tab:list2} summarizes the properties of observed GRBs by MAGIC according to the
informations distributed through the GCN Circular Service. 

\begin{table}[htb]
\begin{center}
\begin{tabular}{r|l|r|r|r|c|c}
   & Burst  & Trigger \# & Energy Range & $T_{90} $ & Fluence & $z$ \\
\hline
1. & GRB~050421   & 115135 & 15-350 keV & 10 s & $1.8\times 10^{-7}$ & -       \\
2. & GRB~050502a   & 2484 & 20-200 keV & 20 s & $1.4\times 10^{-6}$ & 3.79  \\
3. & GRB~050505     & 117504 & 15-350 keV & 60 s & $4.1\times 10^{-6}$ & 4.27 \\
4. & GRB~050509a  & 118707 & 15-350 keV & 13 s & $4.6\times 10^{-7}$ & - \\
5. & GRB~050713a & 145675 & 15-350 keV & 70 s & $9.1\times 10^{-6}$ & - \\
6. & GRB~050904    & 153514 & 15-350 keV & 225 s & $5.4\times 10^{-6}$ & 6.29 \\
7. & GRB~060121   &  4010 & 0.02-1 MeV & 2 s & $4.7\times 10^{-6}$ & - \\
8. & GRB~060203  & 180151 & 15-350 keV & 60 s & $8.5\times 10^{-7}$ & - \\
9. & GRB~060206  & 180455 & 15-350 keV & 11 s & $8.4\times 10^{-7}$ & 4.05 \\
\end{tabular}\\
\end{center}
\caption{\label{tab:list2} Main properties of GRBs observed by MAGIC.
The third column shows the typical energy range of detector on board of the satellite,
while in the fourth, fifth and sixth column there are the corresponding measured duration $T_{90}$,
fluence in [erg~cm$^{-2}$] and redshift.}
\end{table}

\begin{figure}[h]
\epsscale{0.8}
\plotone{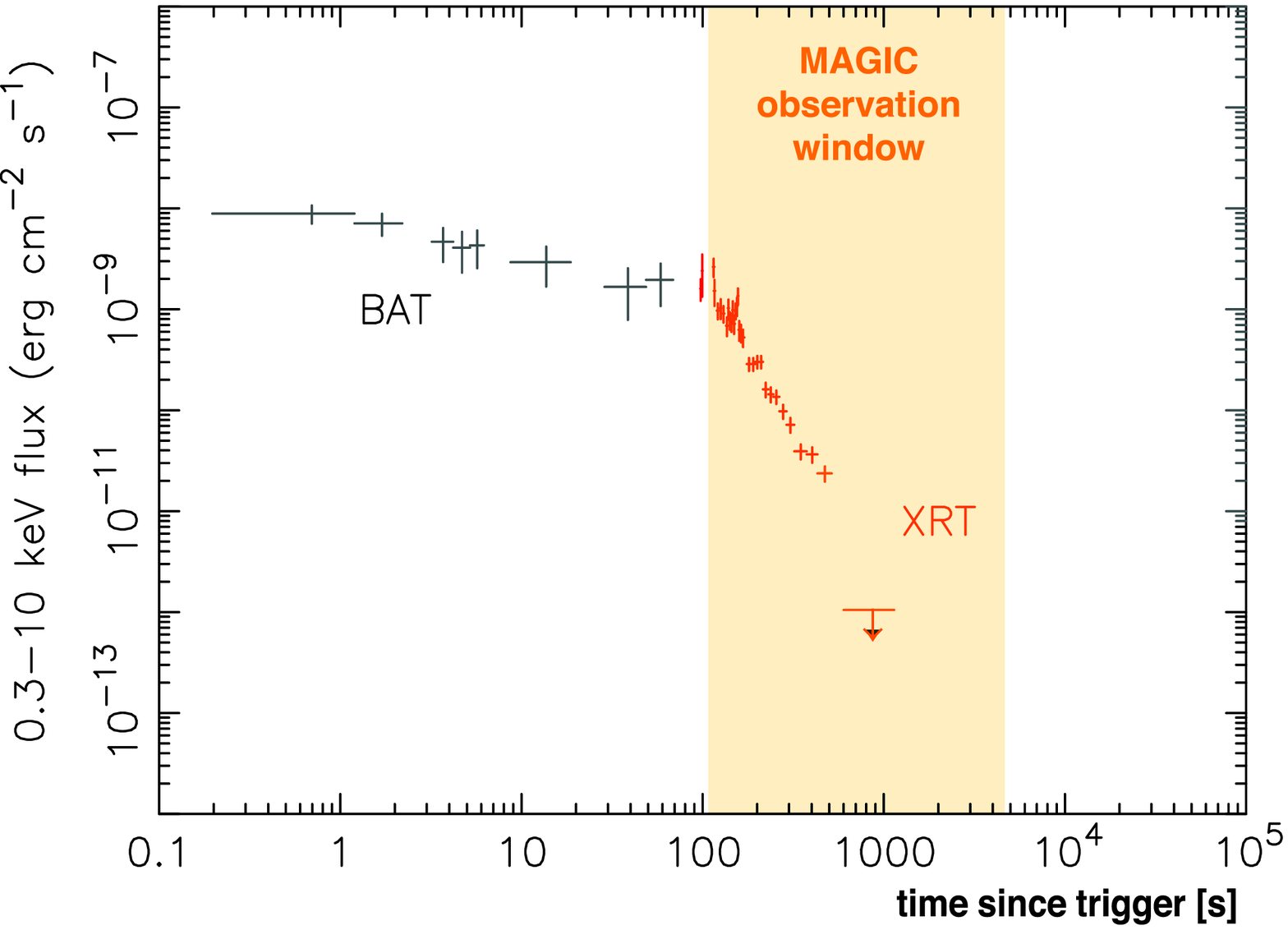}
\caption{Flux of GRB~050421 measured by BAT and XRT. The orange shadowed area represents the MAGIC observation time window and the overlap with Swift data.}
\label{fig:GRB050421BATXRT}
\end{figure}

GRB~050421 was detected by the Burst Alert Telescope (BAT) on board Swift~\cite{GRB050421}.
The other telescope on board Swift, the X-ray Telescope (XRT), could observe
the burst in the 0.2-10~keV range since $T_0 + 97$~s and could detect two
X-ray flares at $T_0 + 110$~s and $T_0 + 154$~s \citep{GODET2006}. 
Figure~\ref{fig:GRB050421BATXRT} 
shows the X-ray light curve of this burst. It can be seen that the MAGIC observation window
is overlapped with the XRT ones on the X-ray afterglow. In particular, the two small X-ray flares
are in the observation window of MAGIC.
No optical counterpart was observed, thus GRB~050421 has been catalogued as a dark burst.

\begin{figure}[h]
\plotone{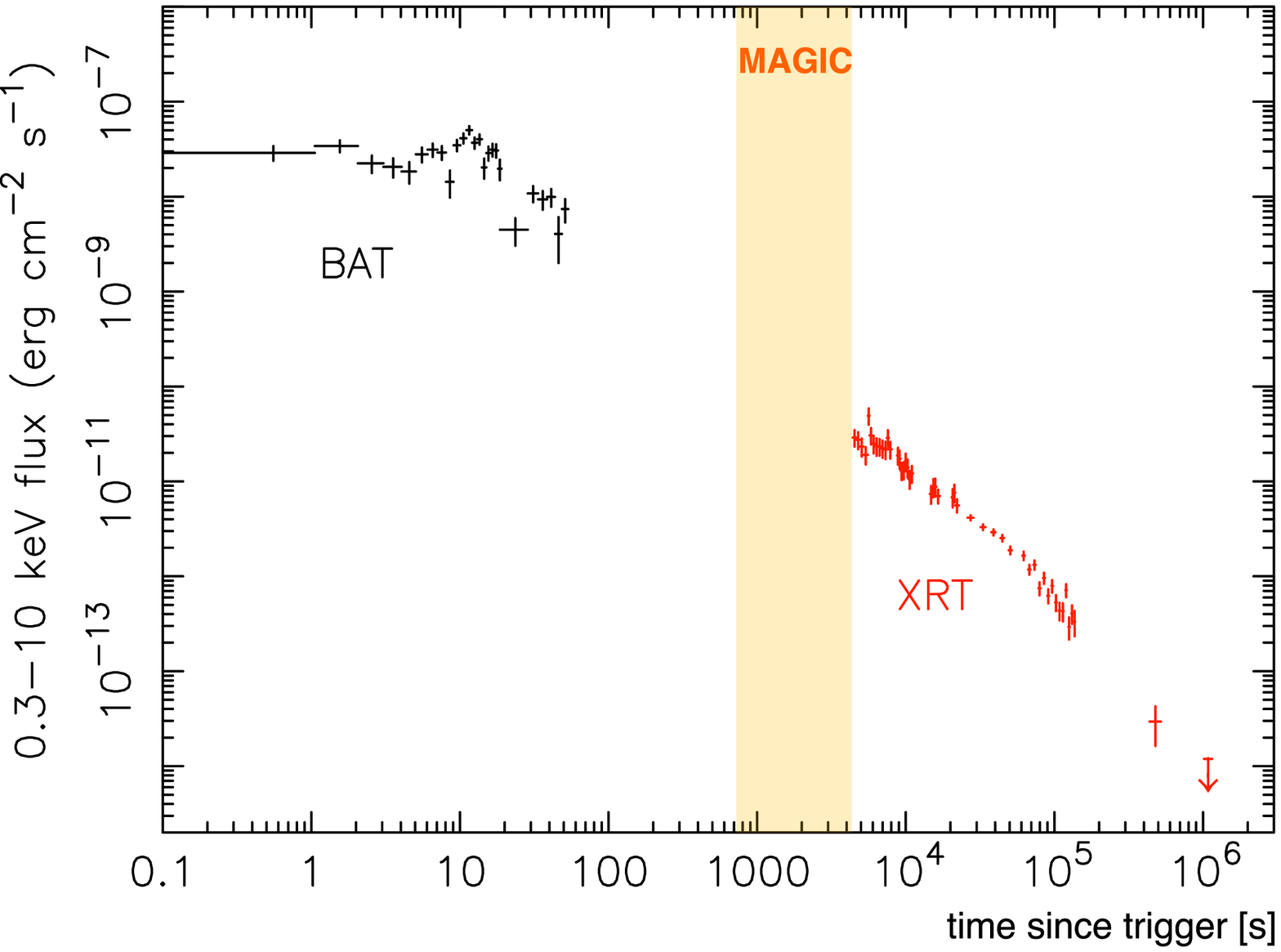}
\caption{\label{fig:GRB050505BATXRT}
Flux of GRB~050505 measured by BAT and XRT. 
The orange shadowed area represents the MAGIC observation time window, starting 717~s after the burst onset.}
\end{figure}

GRB~050502a was triggered by Integral, no X-ray counterpart was observed, but an optical
afterglow followed the burst \citep{GRB050502a, GRB050502a-2, GRB050502a-3}. 
GRB~050505 was triggered by
BAT and its light curve presented three short spikes at $T_{0}$+23.3~s, $T_{0}$+30.4~s and $T_{0}$+50.4s \citep{GRB050505, GRB050505-3}. Both X-ray and optical observations followed the burst, but there was no simultaneous
observation by MAGIC and the other instruments on board the satellite, as shown in 
figure~\ref{fig:GRB050505BATXRT}.

\begin{figure}[h]
\plotone{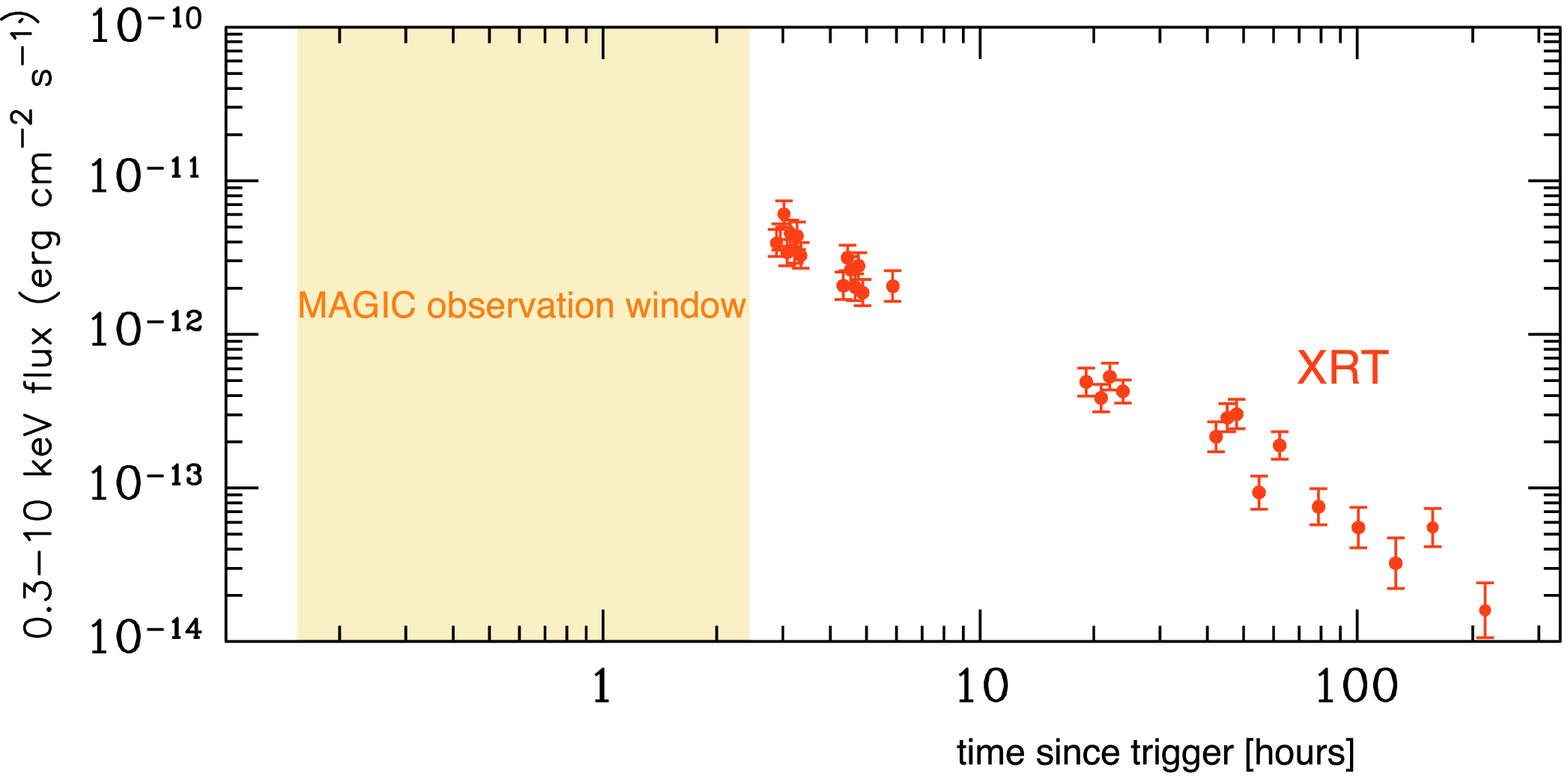}
\caption{Flux of the afterglow of GRB~060121 measured by Swift XRT detector. MAGIC observation window is shown as orange area.}
\label{fig:GRB060121HETE}
\end{figure}

GRB~050509a was triggered by BAT and later XRT could detect an X-ray counterpart 
\citep{GRB050509a, GRB050509a-2}. GRB~060121 is the only short burst observed by MAGIC
and was triggered by HETE-II \citep{GRB060121, GRB060121-2}. XRT observed the afterglow
as well  and detected a fading  X-ray source inside the HETE~II error box \citep{GRB060121-3},
as shown in figure~\ref{fig:GRB060121HETE}. Unfortunately, there is no overlap between MAGIC
observation and XRT observation. 
An optical counterpart was not confirmed by TNG, which did detect, however, a weak source inside the 
XRT error box \citep{GRB060121-4}. Moreover, HST gave no evidence of an optical afterglow, 
although the burst lay close to a faint red galaxy at high redshift \citep{GRB060121-5}. 

\begin{figure}[h]
\plotone{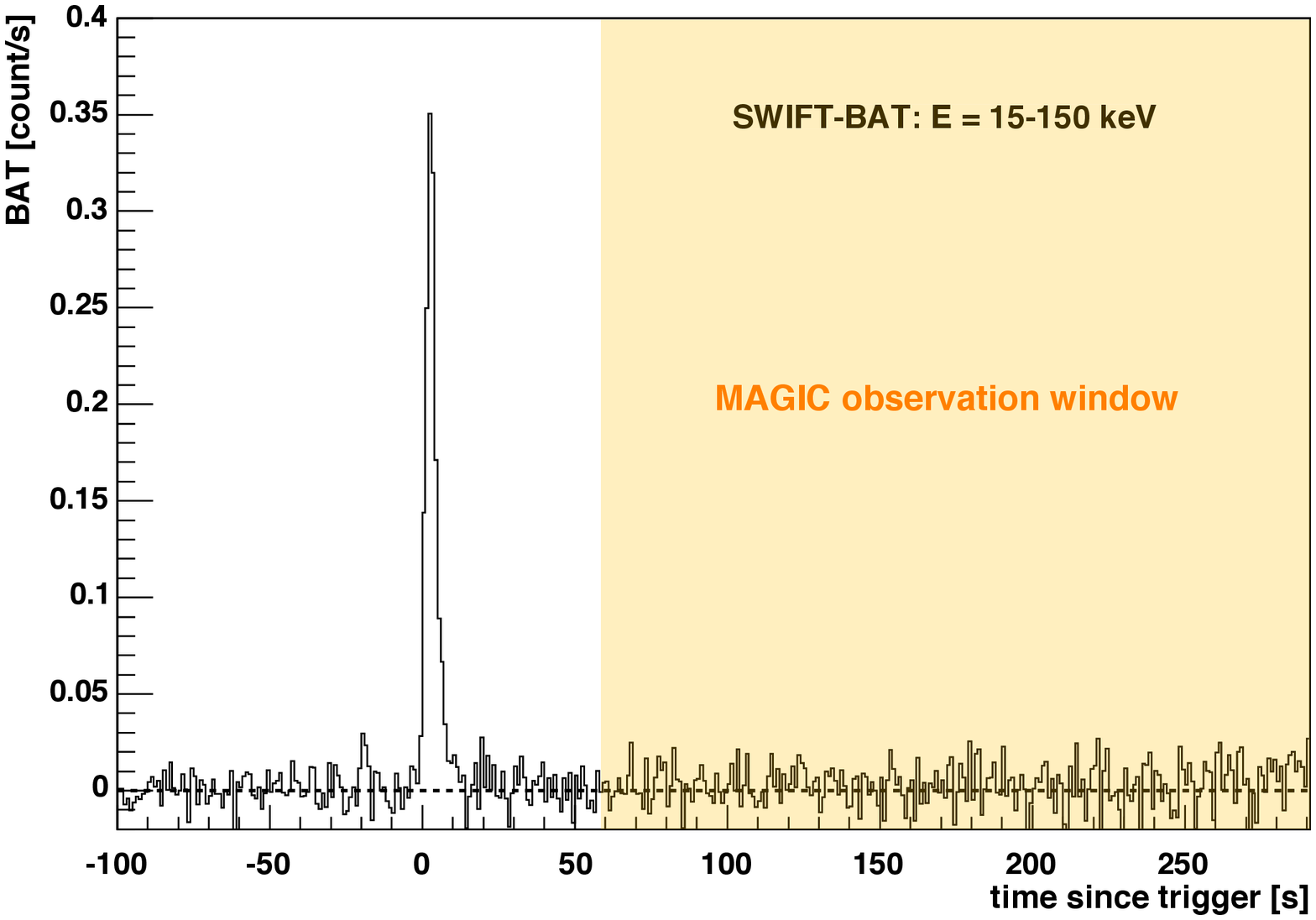}
\caption{Light curve of GRB~060206 measured by BAT, sampled in bins of 1~s. 
The beginning of MAGIC observation is shown shadowed.}
\label{fig:GRB060206BATXRT}
\end{figure}

The last two bursts, GRB~060203 and GRB~060206, are both long bursts triggered by BAT
\citep{GRB060203, GRB060206}. MAGIC data
overlap with XRT data on the X-ray afterglow of GRB~060206, immediately after
BAT data as shown in figure~\ref{fig:GRB060206BATXRT}. No evidence
of flares or of the jet break have been claimed, but optical observations
provided a high redshift value \citep{GRB060206-2, GRB060206-3}.

\subsection{GRB~050713a and GRB~050904, prompt emission observation}

\begin{figure}[h]
\plotone{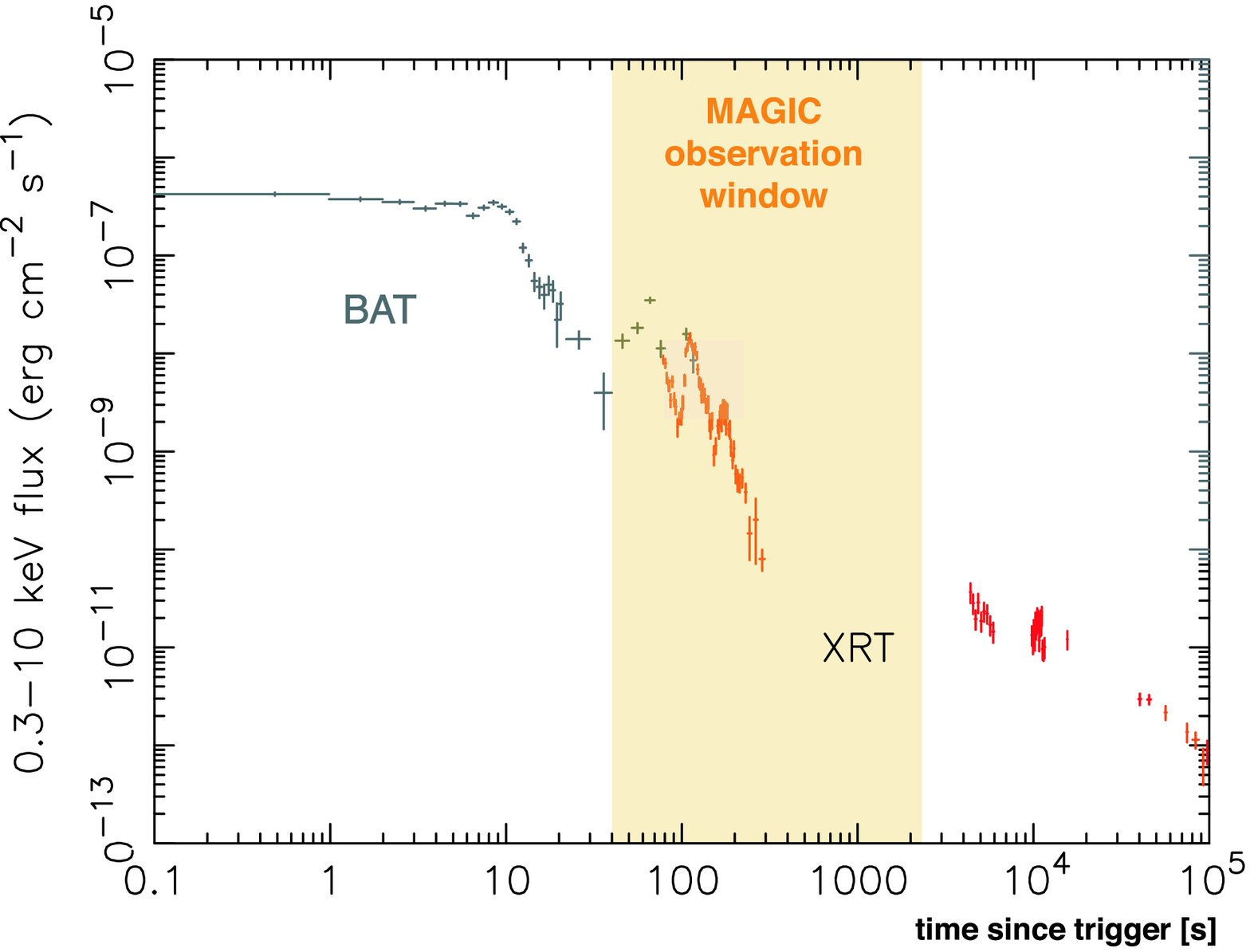}
\caption{\label{fig:GRB050713aBATXRT}
Flux of the prompt and afterglow emission of GRB~050713a measured by BAT and XRT. 
The orange shadowed area illustrates the observation time window of MAGIC.}
\end{figure}

GRB~050713a is of particular interest, being the first burst observed by MAGIC during its prompt emission \citep{albert06}.
On July 13$^\mathrm{th}$, 2005 at 04:29:02~UT the BAT instrument detected a burst 
located at \mbox{R.A. = $21^\mathrm{h} \: 22^\mathrm{m} \: 09^{\,}\fs53$}, 
\mbox{decl. = $+77\degr \: 04\arcmin \: 29^{\,}\farcs50\pm3\arcmin$}   \citep{GRB050713a}.
The MAGIC alert system received and validated the alert 13~s after the burst,
data taking started 40~s after the burst original time $T_0$~\citep{MAGIC_GCN}.
The burst was classified as a bright burst by Swift  with a duration of $T_{90}=70\pm10$~s.
The brightest part of the keV emission occured within $T_0+20$~s,
three smaller peaks followed at $T_0+50$~s, $T_0+65$~s and $T_0+105$~s,
while a \emph{pre-burst\/} peak took place at $T_0-60$~s (see figure~\ref{fig:GRB050713aBATXRT}).
The spectrum, over the interval from $T_0-70$~s to $T_0+121$~s, can be fitted
with a power law with photon index $-1.58\pm0.07$ and yields a fluence of
$9.1\times 10^{-6}\:\mathrm{erg}\;\mathrm{cm}^{-2}$
in the 15-350~keV range~\cite{GRB050713a-2}. 
The burst triggered also the Konus-Wind satellite \cite{GRB050713a-3}, which measured the spectrum of the burst 
during the first 16~s, that is the duration of the first big peak as 
reported by Swift.
In the local coordinate system of MAGIC, GRB~050713a was located at an azimuth angle 
of $-6\degr$ (near North) and a zenith angle of $50\degr$.  The sky region of
the burst was observed during $37\:\mathrm{min}$, until twilight.

\begin{figure}[h]
\plotone{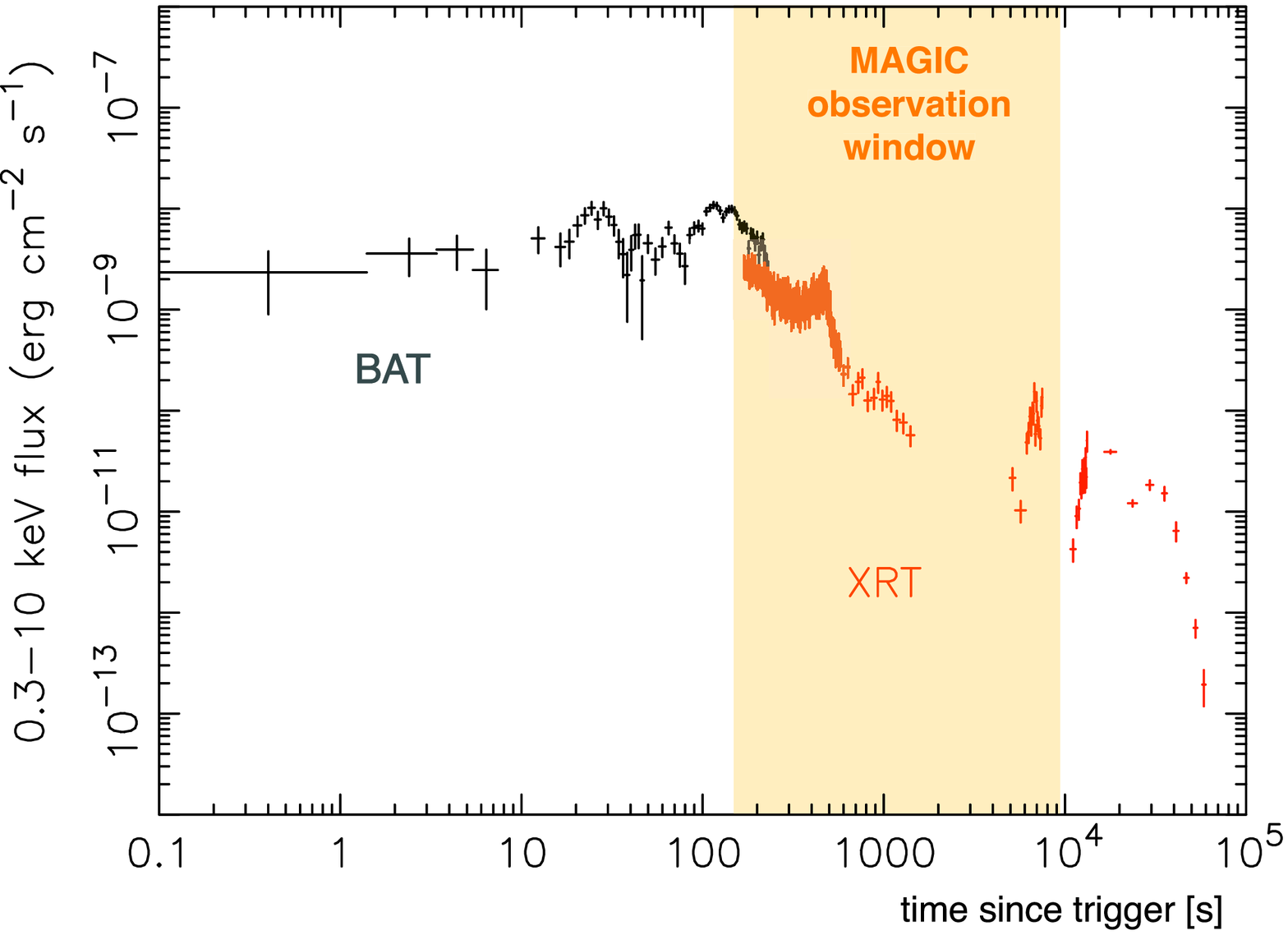}
\caption{Flux measured by BAT and XRT in the prompt and afterglow emission of
GRB~050904. The MAGIC observation window is shown shadowed.}
\label{fig:GRB050904BATXRT}
\end{figure}

Also GRB~050904 is of particular interest, being the second and the latest burst with prompt
emission observed by MAGIC. It was triggered at
01:51:44~UT by BAT, coordinates were 
\mbox{R.A. = $0^\mathrm{h} \: 54^\mathrm{m} \: 50^\mathrm{s}.79$}, \mbox{decl. = $+14^\circ \:05' \: 09''.42\pm3' $}
\citep{GRB050904}.
XRT slewed promptly and started the observation at \mbox{$T_0 + 161$ s}, revealing an
uncatalogued fading source. It is a long burst (\mbox{$T_{90} = 225$ s}), with a total fluence
of $5.4\times 10^{-6}$ erg~cm$^{-2}$ in the \mbox{15-150 keV} range \citep{GRB050904-2}. 
This burst is the most distant burst ever observed,
with an estimated redshift $z=6.29$ \citep{Kawai}.
Its X-ray light curve (see figure~\ref{fig:GRB050904BATXRT}) shows 
a clear X-ray flare at $T_0 + 466$~s \citep{GRB050904-3}, thus in the MAGIC observation window.

\section{Results}

All nine GRBs were analyzed using the MAGIC standard analysis described above.
In this work the image cleaning algorithm using also arrival time information
was used, as it is more robust.
For each GRB a dedicated OFF-source
data set was selected on the basis of being compatible with the
ON-data with respect to several parameters:
zenith angle, as the effective area depends strongly on it;
local brightness of the sky, depending mostly on Moon phase and zenith angle;
trigger rate, depending mostly on atmospheric transparency and
on hardware settings. Loose preliminary cuts were used to remove
unphysical events. After training of the RF,
for each burst a \emph{hadronness} cut was applied which could grant about 90\% 
efficiency on
$\gamma$-ray events according to the corresponding Monte Carlo,
in order to keep high statistics of possible $\gamma$-rays events.

\begin{figure}[htp]
\includegraphics[angle=-90, width=0.9\textwidth]{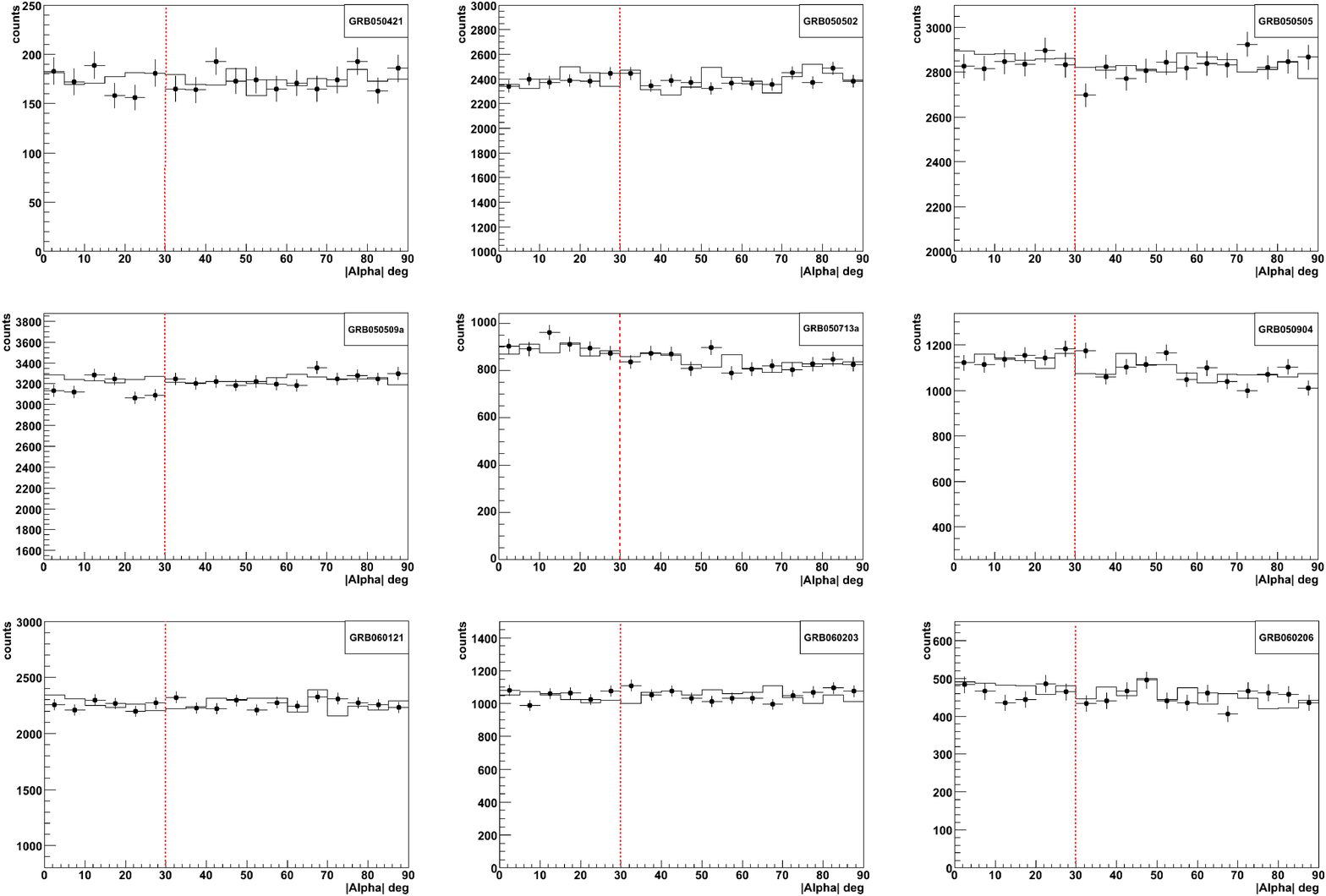}
\caption{Alpha plots of all nine GRBs for the complete data set of each burst. Points with
error bars refer to the burst data, the line refers to the background.}
\label{fig:AlphaPlots}
\end{figure}

The analysis showed no evident signal excess, as can be seen in 
figure~\ref{fig:AlphaPlots}. 
For each burst the \emph{alpha} plot over the whole data set and for 
reconstructed energies greater than 100~GeV is shown. 
The \emph{alpha} distributions of the GRB data sets are flat, as expected 
from background hadronic events, and are 
compatible with the corresponding OFF-source data set. No excess in the signal region, 
i.e. for $\mathrm{\emph{alpha}}<30^\circ$, can be seen.

\begin{figure}[htp]
\includegraphics[angle=-90, width=0.9\textwidth]{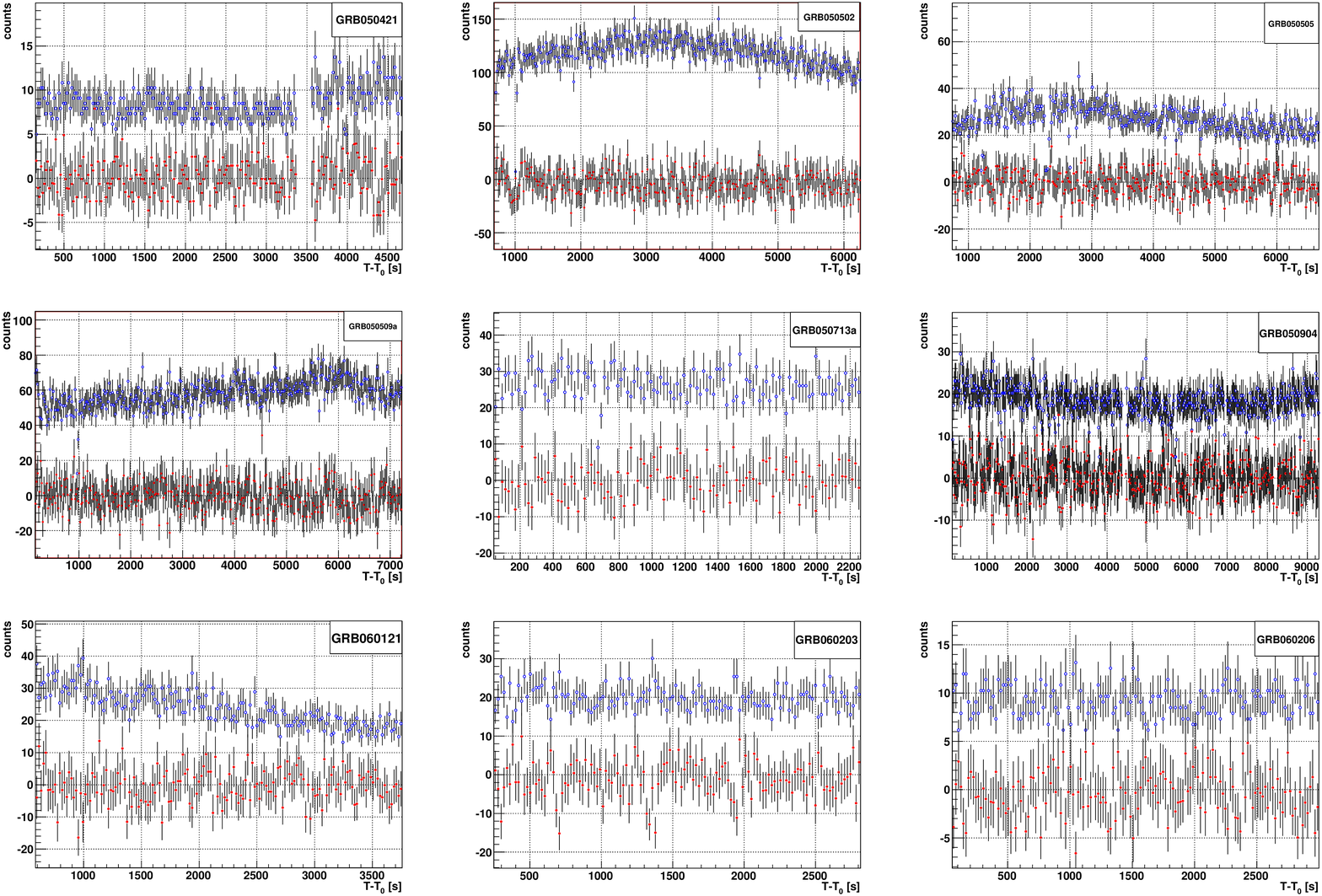}
\caption{Light curves of all nine GRBs for the complete data set of each burst. The background rate
of GRB~050502 is particularly high because of a higher night sky background due to the
Moon light.}
\label{fig:LightCurves}
\end{figure}

No excess was also evidenced using a temporal analysis:
The entire data taking interval was divided into 20-second time bins
and the number of potential $\gamma$-ray events, extracted from the alpha distribution;
they are shown in the light curves of Figure~\ref{fig:LightCurves}:
red filled circles denote
the excess events, blue open circles the background events (offset by 5 from 
excess counts in order to make the plot more readable). 
The distributions of excess events remain zero on average during the observation, 
and no significant variation of the sample 
average is visible with time.

\begin{figure}[htp]
\includegraphics[width=0.9\textwidth]{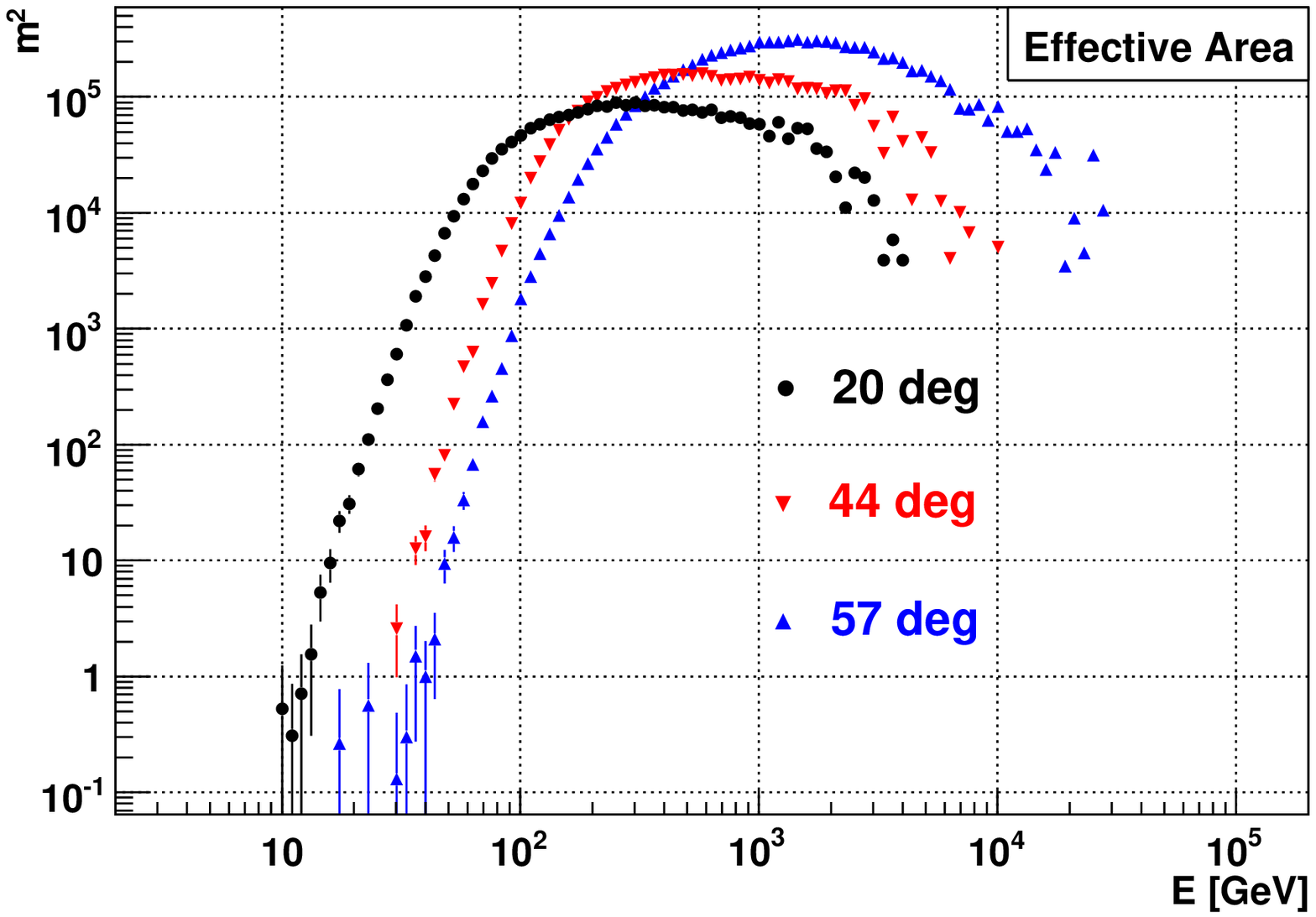}
\caption{Effective area of MAGIC, after typical cuts used in this analysis, for three different zenith
angles.}
\label{fig:aeff}
\end{figure}

Upper limits have been derived for the first 30 minutes of each burst using the Rolke
approach~\cite{Rolke} in six reconstructed energy bins: 80-120~GeV, 120-175~GeV, 
175-225~GeV, 225-300~GeV,
300-400~GeV and 400-1000~GeV.
A systematic uncertainty of 30\% on the efficiency 
has been considered in the upper limit calculation. 
In every reconstructed energy bin the upper limit in number of excess, calculated with the Rolke
approach, has been
converted in flux unit using the effective area, as explained in
appendix~\ref{app:ul}. The typical effective area of MAGIC for different zenith angle is shown
in figure~\ref{fig:aeff}.
Table~\ref{tab:ul} summarizes the upper limits derived for all nine GRBs during the first 30 minutes of data taking. 

\begin{table}[htbp]
\begin{center}
\begin{tabular}{l|c|c|c@{\quad}c@{\quad}c}
{} & \textbf{Energy bin} & \textbf{Energy} & \multicolumn{3}{c}{\textbf{Fluence Upper Limit}} \\
{} & \textbf{[GeV]} & \textbf{[GeV]} & \textbf{[cm$^{-2}$ keV$^{-1}$]} & \textbf{[erg cm$^{-2}$]} & \textbf{C.U.} \\
\hline
\hline
{} & 175-225 & 212.5 & $5.26\times 10^{-16}$ & $3.80\times 10^{-8}$ & 0.20 \\
\cline{2-6}
{} & 225-300 & 275.8 & $3.64\times 10^{-16}$ & $4.43\times 10^{-8}$ & 0.27 \\
\cline{2-6}
\raisebox{1.5ex}[-1.5ex]{\textbf{GRB~050421}} & 300-400 & 366.4 & $5.21\times 10^{-17}$ & $1.12\times 10^{-8}$ & 0.08 \\
\cline{2-6}
{} & 400-1000 & 658.7 & $2.07\times 10^{-17}$ & $1.41\times 10^{-8}$ & 0.14 \\
\hline
\hline
{} & 120-175 & 152.3 & $1.67\times 10^{-15}$ & $6.21\times 10^{-8}$ & 0.27 \\
\cline{2-6}
{} & 175-225 & 219.3 & $2.83\times 10^{-15}$ & $2.18\times 10^{-7}$ & 1.15 \\
\cline{2-6}
\textbf{GRB~050502} & 225-300 & 275.8 & $1.13\times 10^{-15}$ & $1.37\times 10^{-7}$ & 0.83 \\
\cline{2-6}
{} & 300-400 & 360.8 & $7.57\times 10^{-17}$ & $1.58\times 10^{-8}$ & 0.11 \\
\cline{2-6}
{} & 400-1000 & 629.1 & $5.62\times 10^{-17}$ & $3.56\times 10^{-8}$ & 0.35 \\
\hline
\hline
{} & 175-225 & 212.9  & $2.03\times 10^{-15}$  & $1.48\times 10^{-7}$ & 0.76  \\
\cline{2-6}
{} & 225-300 & 275.1 & $2.66\times 10^{-15}$ & $3.22\times 10^{-7}$ & 1.94 \\
\cline{2-6}
\raisebox{1.5ex}[-1.5ex]{\textbf{GRB~050505}} & 300-400 & 363.6 & $5.28\times 10^{-16}$ & $1.11\times 10^{-7}$ & 0.79 \\
\cline{2-6}
{} & 400-1000 & 704.1 & $1.85\times 10^{-17}$ & $1.46\times 10^{-8}$ & 0.15 \\
\hline
\hline
{} & 175-225 & 215.1 & $1.04\times 10^{-15}$ & $7.69\times 10^{-8}$ & 0.40 \\
\cline{2-6}
{} & 225-300 & 273.4 & $1.39\times 10^{-15}$ & $1.67\times 10^{-7}$ & 1.00 \\
\cline{2-6}
\raisebox{1.5ex}[-1.5ex]{\textbf{GRB~050509a}} & 300-400 & 362.8 & $7.74\times 10^{-16}$ & $1.63\times 10^{-7}$ & 1.15\\
\cline{2-6}
{} & 400-1000 & 668.5 & $1.69\times 10^{-16}$ & $1.21\times 10^{-7}$ & 1.22\\
\hline
\hline
{} & 120-175 & 169.9 & $3.63\times 10^{-15}$ & $1.68\times 10^{-7}$ & 0.76 \\
\cline{2-6}
{} & 175-225 & 212.5 & $1.12\times 10^{-15}$ & $8.08\times 10^{-8}$ & 0.42 \\
\cline{2-6}
\textbf{GRB~050713a} & 225-300 & 275.8 & $2.07\times 10^{-15}$ & $2.52\times 10^{-7}$ & 1.52 \\
\cline{2-6}
{} & 300-400 & 366.4 & $3.33\times 10^{-16}$ & $7.16\times 10^{-8}$ & 0.51 \\
\cline{2-6}
{} & 400-1000 & 658.7 & $2.24\times 10^{-17}$ & $1.55\times 10^{-8}$ & 0.15 \\
\hline
\hline
{} & 80-120 & 85.5 & $9.06\times 10^{-15}$ & $1.06\times 10^{-7}$ & 0.32 \\
\cline{2-6}
{} & 120-175 & 140.1 & $3.00\times 10^{-15}$ & $9.42\times 10^{-8}$ & 0.38 \\
\cline{2-6}
{} & 175-225 & 209.9 & $2.18\times 10^{-15}$ & $1.53\times 10^{-7}$ & 0.79 \\
\cline{2-6}
\raisebox{1.5ex}[-1.5ex]{\textbf{GRB~050904}} & 225-300 & 268.9 & $5.82\times 10^{-16}$ & $6.74\times 10^{-8}$ & 0.40 \\
\cline{2-6}
{} & 300-400 & 355.2 & $5.01\times 10^{-16}$ & $1.11\times 10^{-7}$ & 0.71 \\
\cline{2-6}
{} & 400-1000 & 614.9 & $1.26\times 10^{-16}$ & $7.63\times 10^{-8}$ & 0.73 \\
\hline
\hline
\end{tabular}
\end{center}
\end{table}

\begin{table}[htbp]
\begin{center}
\begin{tabular}{l|c|c|c@{\quad}c@{\quad}c}
{} & \textbf{Energy bin} & \textbf{Energy} & \multicolumn{3}{c}{\textbf{Fluence Upper Limit}} \\
{} & \textbf{[GeV]} & \textbf{[GeV]} & \textbf{[cm$^{-2}$ keV$^{-1}$]} & \textbf{[erg cm$^{-2}$]} & \textbf{C.U.} \\
\hline
\hline
{} & 120-175 & 151.3 & $2.64\times 10^{-15}$ & $9.67\times 10^{-8}$ & 0.41 \\
\cline{2-6}
{} & 175-225 & 212.8 & $6.57\times 10^{-16}$ & $4.76\times 10^{-8}$ & 0.25 \\
\cline{2-6}
\textbf{GRB~060121} & 225-300 & 273.7 & $2.13\times 10^{-16}$ & $2.56\times 10^{-8}$ & 0.15 \\
\cline{2-6}
{} & 300-400 & 367.7 & $4.47\times 10^{-16}$ & $9.66\times 10^{-8}$ & 0.69 \\
\cline{2-6}
{} & 400-1000 & 636.4 & $4.84\times 10^{-17}$ & $3.14\times 10^{-8}$ & 0.31 \\
\hline
\hline
{} & 120-175 & 151.5 & $1.10\times 10^{-14}$ & $4.03\times 10^{-7}$ & 1.71 \\
\cline{2-6}
{} & 175-225 & 219.5 & $5.07\times 10^{-16}$ & $3.91\times 10^{-8}$ & 0.21 \\
\cline{2-6}
\textbf{GRB~060203} & 225-300 & 274.0 & $1.57\times 10^{-16}$ & $1.88\times 10^{-8}$ & 0.11 \\
\cline{2-6}
{} & 300-400 & 365.3 & $3.54\times 10^{-16}$ & $7.56\times 10^{-8}$ & 0.54\\
\cline{2-6}
{} & 400-1000 & 639.5 & $4.45\times 10^{-17}$ & $2.91\times 10^{-8}$ & 0.29 \\
\hline
\hline
{} & 80-120 & 85.5 & $1.23\times 10^{-14}$ & $1.44\times 10^{-7}$ & 0.44 \\
\cline{2-6}
{} & 120-175 & 139.9 & $9.83\times 10^{-16}$ & $3.08\times 10^{-8}$ & 0.13 \\
\cline{2-6}
{} & 175-225 & 210.3 & $5.50\times 10^{-16}$ & $3.89\times 10^{-8}$ & 0.20 \\
\cline{2-6}
\raisebox{1.5ex}[-1.5ex]{\textbf{GRB~060206}} & 225-300 & 269.2 & $3.65\times 10^{-16}$ & $4.23\times 10^{-8}$ & 0.25 \\
\cline{2-6}
{} & 300-400 & 355.4 & $6.47\times 10^{-16}$ & $1.31\times 10^{-7}$ & 0.91 \\
\cline{2-6}
{} & 400-1000 & 614.0 & $2.88\times 10^{-17}$ & $1.74\times 10^{-8}$ & 0.17 \\
\hline
\hline
\end{tabular}
\caption{Derived fluence upper limits for the first 30 minutes of data of nine Gamma-Ray Bursts.
The first column shows the reconstructed energy bins in which the analysis has been done.
The second column shows the true energy at which the upper limits have been calculated, 
and is the energy giving the average flux upper limit in the reconstructed energy bin. 
The last column shows the upper limit value in Crab Unit.}
\label{tab:ul}
\end{center}
\end{table}

\section{Discussion}

A preliminary estimate of the observability of GRBs by the MAGIC telescope had 
originally been derived using
 the fourth BATSE catalogue \citep{ICRC, tesigalante}. The GRB spectra were extended to GeV energies 
with a simple power law and using the observed high-energy spectral index. 
The extrapolated fluxes
were finally compared to the estimated MAGIC sensitivity. Setting conservative cuts on observation times and significances,
and assuming an energy threshold of 30~GeV, a 5\,$\sigma$-signal rate of $0.5-2$ per year
was obtained for an assumed observation delay between 15 and 60\,s and a 
BATSE trigger rate of $\sim$\,360/year. 
Taking into account the rate of GRBs \cite{GUETTA}, and extrapolating 
GRB spectra, as observed by BATSE, to VHE using an unbroken power law of 
reasonable power index \citep{Preece}, 
it was foreseen that MAGIC could detect about one GRB per year at a $5\sigma$ level.
A maximal redshift up to $z=2$ was considered.

This estimate must be revised: the Swift alert rate is about a factor~2 lower than 
predicted, and it includes bursts even fainter than those observed by BATSE;
also, for these bursts the effective MAGIC energy threshold at analysis
level was higher than the assumed one ($\sim 80$~GeV);
most important, the distribution of bursts detected by Swift has a
much higher median redshift than expected. As a result,
the number of GRBs that MAGIC can detect is now estimated to lie in the
range of 0.2-0.7 per year.
This number can be expected to increase again with the launch of the GLAST
and AGILE satellites, and the increased number of alerts due to the GRB monitoring by
GLAST, AGILE and Swift altogether.

\section{Conclusions}

MAGIC was able to observe part of the prompt and the early afterglow emission phase of many GRBs 
as a response to the alert system provided by several satellites.  No excess events 
above $\sim 100$~GeV 
were detected, neither during the prompt emission phase nor during the early afterglow. 
We have derived upper limits for the $\gamma$-ray flux between 85 and 1000~GeV.
These limits are compatible with a naive extension of the
power law spectrum, when the redshift is known, up to hundreds of GeV.

For the first time an Atmospheric Cherenkov telescope was able to perform 
direct rapid observations of the prompt emission phase of GRBs. 
This is particulary relevant in the so called ``Swift era''.  
Although strong absorption of the 
high-energy $\gamma$-ray flux by the MRF is expected at high redshifts, 
given its sensitivity to low fluxes and its fast slewing capabilities, 
the MAGIC telescope is currently expected to 
detect about 0.5 GRBs per year, if the GRB spectra extend to 
the energy domain of hundreds of GeV,
following a power law with reasonable indices.

\acknowledgments
\section*{Acknowledgments}
The construction of the MAGIC Telescope was mainly made possible
by the support of the German BMBF and MPG,
the Italian INFN, and the Spanish CICYT, to whom goes our grateful
acknowledgement.
We would also like to thank the IAC for the excellent working
conditions at the Observatorio del Roque de los Muchachos in La Palma.
This work was further supported by ETH Research Grant TH~34/04~3
and the Polish MNiI Grant 1P03D01028.

Facilities: \facility{MAGIC}

\appendix

\section{Upper Limit Calculation}\label{app:ul}

The recorded number of events in a particular reconstructed energy bin 
$\Delta E_\mathrm{rec}$ is:

\begin{equation}\label{eq:excess}
N_{\Delta E_\mathrm{rec}}=\int_0^\infty \phi(E)A_\mathrm{eff}(E|\Delta E_\mathrm{rec}) \: \mathrm{d}E \times \Delta T
\end{equation}

\noindent where $\phi(E)$ is the flux (ph cm$^{-2}$ s$^{-1}$ GeV$^{-1}$), 
$A_\mathrm{eff}(E|\Delta E_\mathrm{rec})$ is
the effective area after all cuts, included the reconstructed energy cut $\Delta E_\mathrm{rec}$, 
and $\Delta T$ the total time interval of observation. It should be noted that the flux $\phi$ and the effective area $A$ depend on the \emph{true} energy, while
the cuts for the selection of the excess
events $N_{\Delta E_\mathrm{rec}}$ and of the effective area $A_\mathrm{eff}(E|\Delta E_\mathrm{rec})$ depend on the reconstructed (estimated) energy. 
The integral is computed in \emph{true} energy $\mathrm{d}E$.

Being the effective area $A_\mathrm{eff}(E|\Delta E_\mathrm{rec})$  depending on energy,
we must assume a spectral shape, in our case the typical power law of a GRB:

\begin{equation}\label{eq:flux}
\phi(E) = k\cdot \left(\frac{E}{E_0}\right)^\beta
\end{equation}

\noindent where $k$ is a normalization factor (ph cm$^{-2}$ s$^{-1}$ GeV$^{-1}$), $\beta$ is 
the average high energy power-law index \mbox{$\beta=-2.5$}
and $E_0$ the pivot energy (e.g. 1 GeV). From equation 
(\ref{eq:excess}) we obtain the normalization factor as

\begin{equation}\label{eq:k}
k_* = \frac{N_{\Delta E_\mathrm{rec}}}
{\int_0^\infty A_\mathrm{eff}(E|\Delta E_\mathrm{rec}) (E/E_0)^\beta \: \mathrm{d}E \times \Delta T}
\end{equation}

\noindent In our case $N_{\Delta E_\mathrm{rec}}$ is the upper limit in number of excess events
calculated with Rolke statistics. The flux upper limit is 

\begin{equation}\label{eq:ul}
\phi_\mathrm{UL}(E) = k_* \cdot \left( \frac{E}{E_0} \right)^\beta
\end{equation}

The value of the energy $E$ is chosen in a very simple way. We calculate the energy which gives
the average flux in  the observed reconstructed energy bin:

\begin{equation}\label{eq:ul1}
\langle \phi \rangle_{A_\mathrm{eff}} 
= \frac{\int_0^\infty \phi(E)A_\mathrm{eff}(E|\Delta E_\mathrm{rec}) \: \mathrm{d}E}
{\int_0^\infty A_\mathrm{eff}(E|\Delta E_\mathrm{rec}) \: \mathrm{d}E} =
\frac{N_{\Delta E_\mathrm{rec}}}
{\int_0^\infty A_\mathrm{eff}(E|\Delta E_\mathrm{rec}) \: \mathrm{d}E \times \Delta T}
\end{equation}

\noindent From equations (\ref{eq:excess}) and (\ref{eq:k}) we can write

\begin{eqnarray}\label{eq:ul2}
\langle \phi \rangle_{A_\mathrm{eff}} & = & 
\frac{N_{\Delta E_\mathrm{rec}}}
{\int_0^\infty A_\mathrm{eff}(E|\Delta E_\mathrm{rec}) (E/E_0)^\beta \: \mathrm{d}E}
\cdot
\frac{\int_0^\infty A_\mathrm{eff}(E|\Delta E_\mathrm{rec}) (E/E_0)^\beta \: \mathrm{d}E}
{\int_0^\infty A_\mathrm{eff}(E|\Delta E_\mathrm{rec}) \: \mathrm{d}E \times \Delta T} \nonumber \\
 {} & {} & {} \nonumber \\
 & = & k_* \cdot \langle \left(E/E_0\right)^\beta \rangle_{A_\mathrm{eff}}
\end{eqnarray}

Defining $(E_*/E_0) \equiv \langle (E/E_0)^\beta \rangle^{1/\beta}_{A_\mathrm{eff}}$,
from equation (\ref{eq:ul}) we can calculate the average flux upper limit
in the reconstructed energy bin:

\begin{equation}\label{eq:ul3}
\phi_\mathrm{UL}(E_*) = k_* \cdot \left( \frac{E_*}{E_0} \right)^\beta 
= \langle \phi_\mathrm{UL} \rangle_{A_\mathrm{eff}}
\end{equation}

Equation \ref{eq:ul3} has been used to calculate the upper limits shown in table~\ref{tab:ul}.

\end{document}